*G. Schmera and L.B. Kish*

# DOUBLING FLUCTUATION-ENHANCED SENSING INFORMATION BY SEPARATING ADSORPTION-DESORPTION AND DIFUSSIVE FLUCTUATIONS[1]

GABOR SCHMERA

*Space and Naval Warfare System Center, Signal Exploitation & Information Management, San Diego, CA 92152-5001, USA*

LASZLO B. KISH

*Texas A&M University, Department of Electrical and Computer Engineering, College Station, TX 77843-3128, USA*

We analyze a (symmetrical) two-sensor arrangement with a joint boundary line between the sensors for fluctuation-enhanced sensing. We show a way to separate the adsorption-desorption signal components from the diffusive signal component. Thus the method generates two independent output spectra which doubles the sensor information for pattern recognition.

Fluctuation-enhanced sensing (FES) to analyze chemical mixtures was proposed [1] to utilize the omnipresence and great sensitivity of low-frequency conductance fluctuations and conductance 1/f noise against structural and environmental changes and inhomogenities/defects in solid state materials [2,3]. In FES we utilize the stochastic signal component due to the statistical interaction between the chemical agent and the sensor material/structure. Doing FES is a complex task which includes not only many aspects of sensor development but also advanced signal processing issues [4-10].

In the present short note, a new method is introduced which is able to distinguish between the adsorption-desorption and diffusive fluctuations in FES devices based on the surface occupancy of sensors by agent molecules [11,12]. This feature results in the doubling of sensor information and higher speed and/or selectivity.

Figure 1 shows the sketch of the two-sensor system to be studied. The adsorbed molecules can diffuse freely and the particles can freely enter from one of the sensor surfaces to that of the other sensor. The space occupied by the two zones may be surrounded by a diffusion barrier which limits the diffusion to these subspaces. If there is no diffusion boundary around the whole system then the particles which leave/enter the system contribute to the adsorption-desorption noise of the given sensor. Other geometries may also be used but they are less simple to fabricate. The time-dependent output signals of the two sensors arte stochastic and defined as follows:

---

*Separating adsorption-desorption and diffusive fluctuation signals*

$$U_1(t) = KN_1(t) \qquad\qquad U_2(t) = KN_2(t) \qquad\qquad (1)$$

where $K$ is a calibration constant and $N_1(t)$ and $N_2(t)$ are the instantaneous numbers of molecules over sensor-1 and sensor-2, respectively.

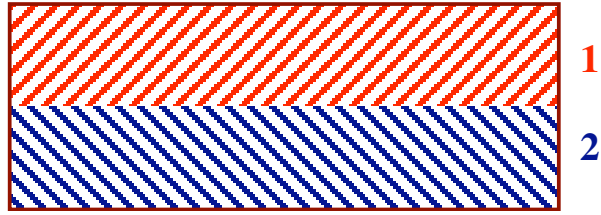

**Figure 1.** Two-sensor arrangement with enhanced joint boundary. Sensor 1 and sensor 2 share an extended joint boundary to enhance cross-correlations of surface diffusion noise. Particles can absorb/desorb over the surface and they execute a random walk (diffusion) over the sensor surface and diffuse over the other sensor, too. If there is no diffusion boundary around the whole system then the particles which leave/enter the system contribute to the adsorption-desorption noise of the given sensor. Other geometries may also be used but they are less simple to fabricate.

The main claims of our study are straightforward and obvious and they are based on adding or subtracting the two sensor outputs. For the sake of simplicity, first let us assume that there is a diffusion boundary around the system and we have two different types of particles; namely one is executing only diffusion (its total number is constant but the number above a given sensor is fluctuation) and the other type is executing only adsorption-desorption (the particles cannot move on the surface however its number is fluctuating). Then:

*i)* The spectrum of $U_1(t) + U_2(t)$ has only absorption-desorption noise. Then the total adsorption-desorption spectrum is:

$$S_{12a} = S^{(+)}(f) = S_{1a}(f) + S_{2a}(f) \quad, \qquad\qquad (2)$$

where $S_{1a}(f)$ and $S_{2a}(f)$ are the adsorption-desorption spectra over *sensor-1* and *sensor-2*, respectively.

*ii)* The spectrum of $U_1(t) - U_2(t)$ is the sum of absorption and diffusion fluctuations:

$$S^{(-)}(f) = S_{1a}(f) + S_{2a}(f) + 4S_{1d}(f) \,, \qquad\qquad (3)$$

where $S_{1d}(f)$ is the diffusion spectrum over *sensor-1* (it is equal to the diffusion spectrum *sensor-2*).

*iii)* After generating the spectra in *i)* and *ii)*, the total diffusion fluctuation spectrum can be obtained by a simple subtraction:

$$S_{12d}(f) = 4S_{1d}(f) = S^{(-)}(f) - S^{(+)}(f) \ . \tag{4}$$

The FES information will be $S_{12a}(f)$ and $S_{12d}(f)$ which are separated adsorption-desorption and diffusion spectra.

Now, let us give up our assumption about two separate particles where one type does adsorption-desorption and the other type does diffusion. Let us assume that we have one type of particle which is doing both processes.

Then Equations 2 and 4 still provide separate information however not the types of particle are separated but the types of processes the particles are executing. Thus the amount of information about the features of the given particle is doubled. This helps to execute a chemical fingerprinting with higher selectivity, speed and sensitivity.

**References**

[1] L.B. Kish, R. Vajtai, C.-G. Granqvist, "Extracting Information from the Noise Spectra of Chemical Sensors: Electronic Nose and Tongue by One Sensor", *Sensors and Actuators B* **71** (2000) 55-59.
[2] M.B. Weissman, *Reviews of Modern Physics* **60** (1988) 537.
[3] L.K.J. Vandamme, "Bulk and surface 1/f noise", *IEEE Transaction on Electron Devices* **36** (1989) 987-992.
[4] L.B. Kish, J. Smulko, P. Heszler, C.G. Granqvist, "On the sensitivity, selectivity, sensory information and optimal size of resistive chemical sensors", *Nanotechnology Perceptions* **3** (2007) 43–52.
[5] Ch. Kwan, G. Schmera, J. Smulko, L.B. Kish, P. Heszler, C.G. Granqvist, "Advanced agent identification at fluctuation-enhanced sensing", *IEEE Sensors*, in press (2008).
[6] G. Schmera, Ch. Kwan, P. Ajayan, R. Vajtai, L.B. Kish, "Fluctuation-enhanced sensing: status and perspectives, *IEEE Sensors*, in press (2008).
[7] S. Gomri, J. L. Seguin, J. Guerin and K. Aguir, *Sensors Actuators B* **114** (2006) 451.
[8] G. H. Huang and S. P. Deng, *Prog. Chem.* **18** (2006) 494.
[9] P. Heszler, R. Ionescu, E. Llobet, L.F. Reyes, J.M. Smulko, L.B. Kish and C.G. Granqvist, "On the selectivity of nanostructured semiconductor gas sensors, *Phys. Stat. Sol. (b)* **244** (2007) 4331-4335.
[10] V.M. Aroutiounian, Z.H. Mkhitaryan, A.A. Shatveryan, F.V. Gasparyan, M. Ghulinyan, L. Pavesi, L.B. Kish and C.G. Granqvis, "Noise spectroscopy of gas sensors", IEEE Sensors, in press (2008).
[11] G. Schmera, L.B. Kish, "Surface diffusion enhanced chemical sensing by surface acoustic waves", *Sensors and Actuators B* **93** (2003) 159–163.
[12] G. Schmera, L.B. Kish, Fluctuation-Enhanced Gas Sensing by Surface Acoustic Wave Devices, *Fluctuation and Noise Letters* **2** (2002) L119-L126.